\documentclass[%
 reprint,
superscriptaddress,
 amsmath,amssymb,
 aps,
]{revtex4-2}

\usepackage{graphicx}
\usepackage{dcolumn}
\usepackage{bm}

\usepackage[colorlinks,
            linkcolor=blue,
            anchorcolor=blue,
            citecolor=blue]{hyperref}

\usepackage{newtxtext}
\usepackage{braket}
\usepackage{subfigure}
\usepackage{makecell}
\usepackage{multirow}
\usepackage{diagbox}
\usepackage{tabularx}
\usepackage[colorlinks,linkcolor=blue]{hyperref}

\usepackage{tikz,xcolor,hyperref}

\definecolor{lime}{HTML}{A6CE39}
\DeclareRobustCommand{\orcidicon}{
\begin{tikzpicture}
\draw[lime, fill=lime] (0,0)
circle[radius=0.16]
node[white]{{\fontfamily{qag}\selectfont \tiny \.{I}D}}; 
\end{tikzpicture}
\hspace{-2mm}
}
\foreach \x in {A, ..., Z}{%
\expandafter\xdef\csname orcid\x\endcsname{\noexpand\href{https://orcid.org/\csname orcidauthor\x\endcsname}{\noexpand\orcidicon}}
}

\begin{document}

\preprint{APS/123-QED}

\title{Real-time parameter estimation for two-qubit systems based on hybrid control}

\author{Yue Tian}
\affiliation{Department of Automation, University of Science and Technology of China, Hefei 230027, China}
\author{Xiujuan Lu}
\affiliation{Department of Mechanical Engineering, University of Hong Kong, Hong Kong 999077, China}
\author{Sen Kuang}
\email{\textcolor{black}{skuang@ustc.edu.cn}}
\affiliation{Department of Automation, University of Science and Technology of China, Hefei 230027, China}
\author{Daoyi Dong}
\affiliation{School of Engineering, Australian National University, Canberra ACT 2601, Australia}
\noaffiliation



\begin{abstract}
In this paper, we consider the real-time parameter estimation problem for a ZZ-coupled system composed of two qubits in the presence of spontaneous emission. To enhance the estimation precision of the coupling coefficient, we first propose two different control schemes, where the first one is feedback control based on quantum-jump detection, and the second one is hybrid control combining Markovian feedback and Hamiltonian control. The simulation results show that compared with free evolution, both control schemes can improve parameter precision and extend system coherence time. Next, on the basis of the two control schemes, we propose a practical single-parameter quantum recovery protocol based on Bayesian estimation theory. In this protocol, by employing batch-style adaptive measurement rules, parameter recovery is conducted to verify the effectiveness of both control schemes.
\end{abstract}

\maketitle


\section{\label{sec:level1}Introduction}

Quantum metrology is a scientific discipline that explores the measurement, estimation, and control of quantum states and quantum systems. As an important subfield, quantum parameter estimation aims at obtaining parameter information by preparing the initial probe state, controlling the evolution process, and seeking the optimal measurements. It primarily focuses on how to leverage the non-classical properties of quantum systems (such as entanglement, superposition, coherence) to surpass the traditional shot-noise limit (SNL) and even achieve the Heisenberg limit (HL)~\cite{1,2,3}. Due to its crucial role in various fields, such as quantum gate calibration in quantum computing~\cite{4,5}, channel estimation in quantum communication~\cite{6,7}, and precision measurement in quantum sensing~\cite{8,9,10}, quantum parameter estimation has become one of the research fields that attracts much attention in the quantum information science community.

Up to the current development of quantum metrology, the majority of literature on parameter estimation focuses on single-qubit systems~\cite{11,12,13,14,15,16,17}, which are relatively simple and easy to control. Remarkable achievements include high-precision measurements of multiplicative Hamiltonian parameters and dissipative coefficients under non-unitary evolution~\cite{3,12}. However, there is a paucity of literature addressing the parameters between multiple qubit systems. In fact, within fields such as quantum communication and quantum computing, the interaction term between two qubits is one of the key resources to generate and control quantum entanglement~\cite{18,19}. Furthermore, understanding the nature of system interactions also contributes to identifying the sources of noise, thereby facilitating improved noise suppression. Hence, accurately estimating the interaction term is crucial for quantum technology applications~\cite{20,21}.

In the field of quantum metrology, control plays a fundamental role in enhancing parameter estimation performance~\cite{22,23,24,25}. In terms of open-loop control, Hamiltonian control based on numerical optimization has been demonstrated to effectively improve the estimation performance of qubit systems, including Gradient ascent pulse engineering (GRAPE)~\cite{14,22}, Krotov’s method~\cite{11}, Deep deterministic policy gradients (DDPG)~\cite{25,26}, Asynchronous Advantage Actor-Critic (A3C)~\cite{27}. In terms of closed-loop control, two commonly used and easily implementable techniques are quantum-jump feedback control and homodyne-mediated feedback control~\cite{11,12,13,28}. These methods make real-time measurements to reduce the disturbance causing by the unpredictable environmental interference. Currently, closed-loop feedback control has been widely utilized for the generation of entangled steady states~\cite{29}, control of entropy uncertainty~\cite{30}, manipulation of quantum discord dynamics in two-atom systems~\cite{31}. This paper considers the problem of estimating the coupling strength in the presence of spontaneous emission noise in the two-qubit system. A hybrid control scheme for the two-qubit system is proposed, which not only retains the compensation function of feedback control against stochastic disturbances but also incorporates the flexibility of Hamiltonian control.

The ultimate goal of the above optimization process is to recover the information of quantum parameters to the maximum extent. Various quantum estimation algorithms have been developed, including Bayesian mean estimation (BME)~\cite{32,33}, Maximum likelihood estimation (MLE)~\cite{34}, and Least squares estimation (LSQ)~\cite{35}. However, the existing literature often treats the optimization process and the parameter recovery step separately, resulting in a loose connection between theoretical schemes and practical protocols for precision limit improvements. This paper provides an overall consideration of the parameter estimation process based on Bayesian estimation theory. By designing adaptive measurement rules and demonstrating the applications of control schemes, we propose a practical single-parameter quantum recovery protocol.

This paper is organized as follows. Section \ref{sec:level2} introduces the Fisher information of parameter estimation and provides a brief description of the physical model. In Section \ref{sec:level3}, the impact of feedback control on parameter estimation precision under different detection efficiencies is investigated, and the hybrid control scheme is designed to further enhance parameter estimation performance. Section \ref{sec:level4} devises adaptive measurement rules and proposes a batch single-parameter quantum recovery protocol based on Bayesian estimation theory. The work is summarized in Section \ref{sec:level5}.

\section{\label{sec:level2}Background Knowledge on Quantum Parameter Estimation}

This section introduces several common precision evaluation metrics and shows how to derive the optimal measurement. Then the evolution model of the two-qubit system under consideration is given, and the representation of feedback control model is further deduced.

\subsection{\label{sec:level2.1}Fisher Information}

In the field of quantum sensing, the core task of parameter estimation is to estimate unknown parameters of systems by manipulating quantum states and employing appropriate measurement operations. In an open quantum system, let $\theta$ represents a single unknown parameter to be estimated, which often encompasses various quantities such as coupling coefficient $g$~\cite{21}, magnetic field strength $\omega$~\cite{11}, dissipation rate $\gamma$~\cite{12}. To begin the estimation process, an initial probe state ${\rho}_0$ is prepared to evolve in a quantum channel $\xi_\theta$. By applying appropriate measurement operators $\left\{\mathrm{M}_y\right\}$ ($\sum_{y}\mathrm{M} _y=I$, where $I$ is the identity matrix) to the evolved state, the probability density function of measurement results $p_\theta\left(y\right)$ can be obtained. Subsequently, with the assistance of suitable algorithms, the effective estimates of the unknown parameters can be calculated~\cite{36}. According to Cramér-Rao bound~\cite{12}, the lower bound on the variance of the unbiased estimator $ \hat{\theta}$ satisfies 
\begin{equation}
\mathrm{Var}\left(\hat{\theta}\right)\geq\frac{1}{n\mathcal{I}_\theta}\geq\frac{1}{{n\mathcal{F}}_\theta}
\label{eq:fir}
\end{equation}
where $\mathrm{Var}(\cdot)$ is the variance operator, $n$ is the number of repeated experiments, $\mathcal{I}_\theta$ represents the classical Fisher information (CFI), $\mathcal{F}_\theta$ represents the quantum Fisher information (QFI)~\cite{36}. From Eq. (\ref{eq:fir}), it is clear that larger Fisher information implies a smaller variance limit for the estimator, thereby enabling a higher level of achievable estimation precision.

In classical parameter estimation theory, CFI is an important statistical concept used to describe the distinguishability of the probability space, thereby quantifying the limits of estimation precision~\cite{37}. For a set of discrete measurement result probability distributions $p_\theta\left(y\right)=\mathrm{Tr}(\rho _{\theta }\mathrm{M}_y )$, $\mathcal{I}_\theta$ can be defined as~\cite{14}
\begin{equation}
\mathcal{I}_\theta=\sum_{y}{p_\theta\left(y\right)\left[\frac{\mathrm{d}{\ln{p}}_\theta\left(y\right)}{\mathrm{d}\theta}\right]^2}
\label{eq:sec}
\end{equation}
where ${\ln{}}(\cdot)$ represents the natural logarithm, and $\mathrm{Tr}(\cdot)$ represents the trace operation. CFI is fundamentally a function of measurements. For achieving higher estimation precision, the choice of measurement operators is crucial. Using the Cauchy-Schwartz inequality~\cite{37}, it can be proved that CFI is equivalent to QFI  under the action of the optimal measurement operators, i.e., $\mathrm{max}_{\left\{\mathrm{M}_y\right\}}\mathcal{I}_\theta\left(\rho_\theta,\left\{\mathrm{M}_y\right\}\right)=\mathcal{F}_\theta$~\cite{38}. 

As a basic quantity in quantum physics, QFI has widespread applications in quantum metrology~\cite{22}, quantum phase transitions~\cite{38}, and other fields. $\mathcal{F}_\theta$ can be defined as~\cite{11}
\begin{equation}
\mathcal{F}_\theta=\mathrm{Tr}\left[\rho_\theta\left(L_s^\theta\right)^2\right]
\label{eq:thi}
\end{equation}
Here, $L_s^\theta$ represents the symmetric logarithmic derivative (SLD) operator of $\theta$, which is similar to the logarithmic derivative (LD) in classical statistics. SLD can capture the response information of quantum states to small changes in parameters, which is also an important tool to determine the optimal measurement and plays a key role in quantum parameter estimation~\cite{39}. SLD can be defined as
\begin{equation}
\partial_\theta\rho_\theta=\frac{1}{2}\left(L_s^\theta\rho_\theta+\rho_\theta L_s^\theta\right)
\label{eq:for}
\end{equation}
where $\partial_\theta\rho_\theta$ is the partial derivative of the quantum state $\rho_\theta$ with respect to $\theta$.

By conducting spectral decomposition on the density matrix $\rho_\theta$, it yields $\rho_\theta=\sum_{i}\mathrm{\lambda}_i\ket{\lambda_i}\bra{\lambda_i}$. Then by substituting it into Eq. (\ref{eq:for}), we can obtain $\braket{\lambda_i|\partial_\theta\rho_\theta|\lambda_j}=\frac{1}{2}\left(\mathrm{\lambda}_i+\mathrm{\lambda_j}\right)\braket{\lambda_i|L_s^\theta|\lambda_j}$. Further solving for $L_s^\theta$, we have
\begin{equation}
L_s^\theta=\sum_{i,j,\mathrm{\lambda}_i+\mathrm{\lambda_j}>0}{2\frac{\braket{\lambda_i|\partial_\theta\rho_\theta|\lambda_j}}{\mathrm{\lambda}_i+\mathrm{\lambda_j}}\ket{\lambda_i}\bra{\lambda_j}}.
\label{eq:fif}
\end{equation}

Since the SLD operator is a Hermitian operator, it can be diagonalized, whose eigenvectors correspond to a set of orthogonal normalized bases. For single-parameter estimation problems, the eigenvectors of SLD can be used to construct a set of optimal measurement bases~\cite{38}. Here, the set of eigenvectors of $L_s^\theta$ is denote as $\{\ket{e_i}\}$, and the corresponding projectors are considered as a set of positive-operator valued measures (POVM), denoted as $E_i=\{\ket{e_i}\bra{e_i}\}$. Then the probability of the $i$th measurement result is $\braket{e_i|\rho_\theta|e_i}$. With $L_s^\theta=\sum_{i}e_i\ket{e_i}\bra{e_i}$ and Eq. (\ref{eq:for}), we derive $\braket{e_i|\partial_\theta\rho_\theta|e_i}=e_i\braket{e_i|\rho_\theta|e_i}$. Substituting this equation into Eqs. (\ref{eq:sec}) and (\ref{eq:thi}), we have
\begin{equation}
\begin{split} 
\mathcal{I}_\theta&\le \sum_{i}\frac{\braket{e_i|\partial_\theta\rho_\theta|e_i}^2}{\braket{e_i|\rho_\theta|e_i}}=\sum_{i}{e_i}^2\braket{e_i|\rho_\theta|e_i} \\
&=\mathrm{Tr}\!\left[\sum_{i}{{e_i}^2\rho_\theta\ket{e_i}\bra{e_i}}\right]\!=\!\mathcal{F}_\theta
\label{eq:six}
\end{split} 
\end{equation}

From Eq. (\ref{eq:six}), it can be proved that measurements under the set of projection measurement bases $E_i$ maximize the estimation precision of parameter~\cite{38}. Note that, in most cases, $E_i$ depends on the true value of $\theta$. Considering the limitations of prior knowledge in practical situations, it usually needs to search for the optimal measurements in an adaptive manner. 

This paper focuses on enhancing the single-parameter estimation precision. In this case, there always exists a specific set of optimal measurements that can achieve the precision defined by QFI~\cite{38}, as expressed in Eq. (\ref{eq:thi}). Therefore, QFI is chosen as the objective function for the optimization process. From Eq. (\ref{eq:thi}) and Eq. (\ref{eq:for}), it is clear that maximizing QFI depends on the evolved state $\rho_\theta$ and its sensitivity to the parameter. This dependency involves factors such as the initial state $\rho_0$, degree of dissipation, and control Hamiltonian~\cite{38}.

\subsection{\label{sec:level2.2}Physical model}

Consider a ZZ-coupled system composed of two spin-1/2 subsystems, with energy levels denoted as $\ket{e^{\left(k\right)}}$ and $\ket{f^{\left(k\right)}}$ for each qubit $(k=1,2)$. This system model has widespread applications in strong-field regimes~\cite{22}. In the absence of control, the dynamics of the system can be described by the Lindblad master equation:
\begin{equation}
\dot{\rho}=\mathcal{L}\left(\rho\right)=-\frac{i}{\hbar}\left[H,\rho\right]+\mathcal{L}_\mathcal{D}\left(\rho\right)
\label{eq:sev}
\end{equation}
where $\mathcal{L}$ represents the superoperator of the system's evolution, and the superoperator $\mathcal{L}_\mathcal{D}\left(\rho\right)$ describes the decoherence phenomenon of the quantum system in the environment. $\left [ \cdot , \cdot \right ]  $ is the commuting operator. $\hbar$ is the reduced Planck constant (this article takes $\hbar=1$). $H=H_0+H_c$, where $H_0$ is the free Hamiltonian and $H_c$ is the control Hamiltonian. The free Hamiltonian of the ZZ-coupled system can be expressed as~\cite{21}
\begin{equation}
H_0=\omega_1\sigma_z^{\left(1\right)}+\omega_2\sigma_z^{\left(2\right)}+g\sigma_z^{\left(1\right)}\sigma_z^{\left(2\right)}
\label{eq:eig}
\end{equation}
where $\sigma_z^{\left(1\right)}=\sigma_z\otimes I$, $\sigma_z^{\left(2\right)}=I\otimes\sigma_z$, and $\sigma_z=\ket{e}\bra{e}-\ket{f}\bra{f}$. $\omega_1$ and $\omega_2$ represent the local frequencies of the two qubits, respectively. $g$ is the interaction strength of the ZZ coupling between the two subsystems, which is the parameter to be estimated. Assuming that it has a true value of $g^\ast$. In the case of local spontaneous emission for each qubit, the dissipative term $\mathcal{L}_\mathcal{D}\left(\rho\right)$ can be described as
\begin{equation}
\mathcal{L}_\mathcal{D}\!\left(\rho\right)\!=\!\!\!\!\sum_{k=1,2}\!\!\!{\gamma_k\!\left[\sigma_-^{\left(k\right)}\rho\sigma_+^{\left(k\right)}\!\!-\!\frac{1}{2}\!\left(\sigma_+^{\left(k\right)}\sigma_-^{\left(k\right)}\rho\!+\!\rho\sigma_+^{\left(k\right)}\sigma_-^{\left(k\right)}\!\right)\!\right]}
\label{eq:nin}
\end{equation} \\
where $\gamma_k$ represents the dissipation rate of the $k$th qubit. In this paper, we assume that the dissipative rates for both qubits are the same, i.e., $\gamma_1=\gamma_2=\gamma$ $(\gamma\geq0)$. $\sigma_-^{\left(k\right)}=\ket{g^{\left(k\right)}}\bra{e^{\left(k\right)}}$ and $\sigma_+  ^{\left(k\right)}=\ket{e^{\left(k\right)}}\bra{g^{\left(k\right)}}$ serve as the lowering and raising operators of the $k$th qubit, respectively. 

Note that the presence of the dissipative term introduces information loss during the evolution of the quantum state, resulting in a decrease in the precision of parameter estimation. To address this challenge, feedback control is often used as a useful tool~\cite{38}. Before introducing system model under feedback control, we discuss the relationship between open quantum dynamics and quantum measurement. By discretizing the stochastic master equation (\ref{eq:sev}) and comparing it with the general form of quantum measurement $\rho\left(t+\mathrm{d}t\right)=\sum_{k=0,1}{\Omega_k\left(\mathrm{d}t\right)\rho\left(t\right)\Omega_k^\dagger\left(\mathrm{d}t\right)}$\textsuperscript{~\cite{40}}, one can get
\begin{equation}
\left\{\begin{matrix}
\begin{aligned}
&\Omega_1\left(\mathrm{d}t\right)=\sqrt{\gamma\mathrm{d}t}\sigma_- \\
&\Omega_0\left(\mathrm{d}t\right)=I-\left(iH+\frac{\gamma}{2}\sigma_+\sigma_-\right)\mathrm{d}t
\end{aligned}
\end{matrix}\right.
\label{eq:ten}
\end{equation} \\
where $\Omega_1\left(\mathrm{d}t\right)$ and $\Omega_0\left(\mathrm{d}t\right)$ represent two different measurement operators. To be specific, when a photon is detected as a result of an energy level transition, the measurement outcome is recorded as 1, and the transition process is represented by the operator $\Omega_1\left(\mathrm{d}t\right)$. On the contrary, when no transition event is detected, the measurement outcome is recorded as 0, it is represented by the operator $\Omega_0\left(\mathrm{d}t\right)$. More details can be found in Ref. \cite{11}.

Now, we consider the dynamic model of the system in the presence of feedback control. Here, the output of system (\ref{eq:sev}) is continuously monitored through a photon detector $D$. When the detector $D$ detects the photons emitted by  the energy level transition, a unitary evolution effect represented by $U_{fb}$ is triggered as feedback on the spin to correct the dynamics of the system. No control is applied at other times, and $U_{fb}$ is precisely the feedback control law that needs to be designed subsequently. The density matrix with feedback control at time $t+\mathrm{d}t$ can be represented as
\begin{equation}
\begin{split} 
\rho\left(t+\mathrm{d}t\right)=&U_{fb}\Omega_1\left(\mathrm{d}t\right)\rho\left(t\right)\Omega_1^\dagger\left(\mathrm{d}t\right)U_{fb}^\dagger+ \\
&\Omega_0\left(\mathrm{d}t\right)\rho\left(t\right)\Omega_0^\dagger\left(\mathrm{d}t\right).
\label{eq:add}
\end{split} 
\end{equation}

Substituting Eq. (\ref{eq:ten}) into Eq.(\ref{eq:add}), we have Eq. (\ref{eq:ele}).
\begin{widetext}
\begin{equation}
\rho\left(t+\mathrm{d}t\right)=\!\rho\left(t\right)+\!\left\{\!-i\left[H,\rho\left(t\right)\right]\!+\!\sum_{k=1,2}{\gamma_k\!\left[U_{fb}\sigma_-^{\left(k\right)}\rho\left(t\right)\sigma_+^{\left(k\right)}U_{fb}^\dagger\!-\!\frac{1}{2}\left(\sigma_+^{\left(k\right)}\sigma_-^{\left(k\right)}\rho\left(t\right)\!+\!\rho\left(t\right)\sigma_+^{\left(k\right)}\sigma_-^{\left(k\right)}\right)\right]}\right\}\mathrm{d}t\!+\!O(\mathrm{d}t^2 ).
\label{eq:ele}
\end{equation}
\begin{equation}
\dot{\rho}=-i\left[H,\rho\right]+\sum_{k=1,2}{\gamma_k\left[U_{fb}\sigma_-^{\left(k\right)}\rho\sigma_+^{\left(k\right)}U_{fb}^\dagger-\frac{1}{2}\left(\sigma_+^{\left(k\right)}\sigma_-^{\left(k\right)}\rho+\rho\sigma_+^{\left(k\right)}\sigma_-^{\left(k\right)}\right)\right]}
\label{eq:twe}
\end{equation}
\end{widetext}

When $\mathrm{d}t$ is small enough, the second-order infinitesimal term $O(\mathrm{d}t^2)$ can be ignored, and thus the dynamics under feedback control can be described as Eq. (\ref{eq:twe}). In fact, Eq. (\ref{eq:twe}) is an ideal feedback model that does not consider detection rate limitations, and we consider a more realistic situation with details in Section \ref{sec:level3}. To preserve the Markovian nature of the system, the feedback control must take effect within a short time after the detection event is triggered.

\section{\label{sec:level3}Enhancing Quantum Parameter Estimation Precision}

As we know, the quantum parameter estimation problem mainly includes four steps: 1) preparation of the initial state; 2) parameterization; 3) measurement; 4) parameter recovery~\cite{25}. In this section, we investigate the first two steps. And two control schemes are proposed to improve parameter estimation precision.

Regarding the preparation of the initial state $\ket{\psi_{opt}(0)}$, the objective is to maximize the sensitivity to small variations in the parameter to be estimated. Since system (\ref{eq:sev}) can achieve the highest Hamiltonian parameter estimation precision only in the absence of noise~\cite{22}, the optimal probe state should be a pure state. Let the probe state at time $t=0$ be $\ket{\psi(0)}=a\ket{00}+b\ket{01}+c\ket{10}+d\ket{11}$, and the quantum state at time $T$ be $\ket{\psi(T)}$. The corresponding QFI calculation formula can be written as follows~\cite{22}:
\begin{equation}
\mathcal{F}_{g  } \!\!=\!4\!\operatorname{Re}\!\left[\left\langle\partial_{g } \psi(T) \!\!\mid\!\! \partial_{g } \psi(T)\right\rangle\!\!-\!\!\left\langle\partial_{g } \psi(T) \!\!\mid\!\! \psi(T)\right\rangle\!\!\left\langle\psi(T) \!\!\mid\!\! \partial_{g } \psi(T)\right\rangle\right]
\label{eq:thith}
\end{equation}
where $\operatorname{Re}[\cdot]$ represents the real part. Note that the parameter $\theta$ in Eq. (\ref{eq:thi}) is specifically expressed as the coupling coefficient $g$, and in the remainder of the paper, we use $g$ instead of $\theta$. $\ket{\partial_{g } \psi(T)}$ is the partial derivative of the quantum state $\ket{\psi\left(T\right)}$ with respect to the coupling coefficient $g$. Using the pure state evolution formula $\ket{\psi\left(T\right)}=\ e^{-iHT}\ket{\psi\left(0\right)}$~\cite{22}, and substituting $\ket{\psi\left(T\right)}=\ e^{-i\left(w_1\sigma_z^{\left(1\right)}+w_2\sigma_z^{\left(2\right)}+g\sigma_z^{\left(1\right)}\sigma_z^{\left(2\right)}\right)T}\ket{\psi\left(0\right)}$ into Eq. (\ref{eq:thith}), we have
\begin{equation}
\mathcal{F}_g=4T^2\left(1{-\left(\braket{\psi(0)|\sigma_z^{\left(1\right)}\sigma_z^{\left(2\right)}|\psi(0)}\right)}^2\right).
\label{eq:forth}
\end{equation}

To achieve the highest sensitivity, its metric $\mathcal{F}_g$ should be maximized. This needs to satisfy the condition $\braket{\psi(0)|\sigma_z^{\left(1\right)}\sigma_z^{\left(2\right)}|\psi(0)}=0$, while simultaneously ensuring the normalization of the state vector, leading to
\begin{equation}
\left\{\begin{matrix}
\left|a\right|^2-\left|b\right|^2-\left|c\right|^2+\left|d\right|^2=0 \\
\left|a\right|^2+\left|b\right|^2+\left|c\right|^2+\left|d\right|^2=1
\end{matrix}\right.
\Rightarrow \left\{\begin{matrix}
 \left|a\right|^2+\left|d\right|^2=\frac{1}{2}\\
\left|b\right|^2+\left|c\right|^2=\frac{1}{2}
\end{matrix}\right..
\label{eq:fifth}
\end{equation}

At this point, we can obtain the optimal probe state as
\begin{equation}
\ket{\psi_{opt}(0)}=\frac{1}{2}\ket{00}+\frac{1}{2}\ket{01}+\frac{1}{2}\ket{10}+\frac{1}{2}\ket{11}=\ket{++}
\label{eq:addtwo}
\end{equation}
where $\ket{+}=\frac{1}{\sqrt2}\left(\ket{0}+\ket{1}\right)$. Unless otherwise specified in subsequent studies, this state is chosen as the initial probe state.

After preparing the initial state, we focus on the optimization of the parameterization. First, QFI under the free evolution is studied. Then we discuss the parameter precision when feedback control is applied. Finally, the impact of hybrid control on parameter estimation performance is further explored.           

\subsection{\label{sec:level3.1}Without control}

First, we analyze the estimation precision under free evolution without control. Considering system (\ref{eq:sev}), we estimate the coupling coefficient $g$ between the qubits. Assuming that the density matrix of the quantum state at time $t$ is
\begin{equation}
\rho\left(t\right)=\left[\begin{matrix}\begin{matrix}\rho_{11}\left(t\right)&\rho_{12}\left(t\right)\\\rho_{12}^\ast\left(t\right)&\rho_{22}\left(t\right)\\\end{matrix}&\begin{matrix}\rho_{13}\left(t\right)&\rho_{14}\left(t\right)\\\rho_{23}\left(t\right)&\rho_{24}\left(t\right)\\\end{matrix}\\\begin{matrix}\rho_{13}^\ast\left(t\right)&\rho_{23}^\ast\left(t\right)\\\rho_{14}^\ast\left(t\right)&\rho_{24}^\ast\left(t\right)\\\end{matrix}&\begin{matrix}\rho_{33}\left(t\right)&\rho_{34}\left(t\right)\\\rho_{34}^\ast\left(t\right)&\rho_{44}\left(t\right)\\\end{matrix}\\\end{matrix}\right]
\label{eq:sixth}
\end{equation}

\begin{figure}
    \centering
    \includegraphics[width=\linewidth, height=6cm]{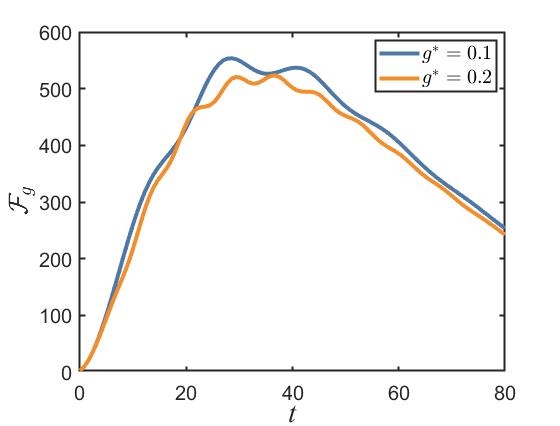}
    \caption{\label{Fig.1} The curves of QFI with time t under different $g^\ast$ without control. The blue solid and the orange solid curves represent $g^\ast=0.1$, and $g^\ast=0.2$, respectively. Other parameters are chosen as $\omega_1=\omega_2=1$, $\gamma=0.05$.}
\end{figure}

By substituting Eq. (\ref{eq:sixth}) into Eq. (\ref{eq:sev}) and taking into account the optimal initial probe state $\ket{\psi_{opt}(0)}$ obtained previously, the analytical solutions for the elements of $\rho\left(t\right)$ can be accurately derived as follows: 
\begin{align}
\rho_{11}(t) &= \frac{1}{4} \exp\left(-(\gamma_1 + \gamma_2)t\right) \nonumber \\
\rho_{12}(t) &= \frac{1}{4}\exp\left(-\left(2\gamma_1+\gamma_2\right)t-i\left(2\omega_2+2g\right)t\right) \nonumber \\
\rho_{13}(t) &= \frac{1}{4}\exp\left(-\left(\gamma_1+2\gamma_2\right)t-i\left(2\omega_1+2g\right)t\right) \nonumber \\
\rho_{14}(t) &= \frac{1}{4}\exp\left(-\frac{1}{2}\left(\gamma_1+\gamma_2\right)t-i\left(2\omega_1+2\omega_2\right)t\right) \nonumber \\
\rho_{22}(t) &= \frac{1}{2}\exp\left(-\gamma_1t\right)-\frac{1}{4}\exp\left(-\left(\gamma_1+\gamma_2\right)t\right) \nonumber \\
\rho_{23}(t) &= \frac{1}{4}\exp\left(-\frac{1}{2}\left(\gamma_1+\gamma_2\right)t-2\left(\omega_1-\omega_2\right)t\right) \nonumber \\
\rho_{24}(t) &= \exp\left(\chi_1\cdot\frac{2g-i\gamma_2}{2\left(4g-i\gamma_2\right)}\right)+ \nonumber \\
&\quad i\gamma_2\cdot \exp\left(-\frac{1}{2}\left(\gamma_1+2\gamma_2\right)t+i\left(4g+4\omega_1\right)t\right)\varrho_2 \nonumber \\
\rho_{33}(t) &= \frac{1}{2}\exp\left(-\gamma_2t\right)-\frac{1/4}\exp\left(-\left(\gamma_1+\gamma_2\right)t\right) \nonumber \\
\rho_{34}(t) &= \exp\left(\chi_2\cdot\frac{2g-i\gamma_1}{2\left(4g-i\gamma_1\right)}\right)+ \nonumber \\
&\quad i\gamma_1\cdot \exp\left(-\frac{1}{2}\left(2\gamma_1+\gamma_2\right)t+i\left(4g+4\omega_2\right)t\right)\varrho_1 \nonumber \\
\rho_{44}(t) &= \frac{1}{4}\exp\left(-\left(\gamma_1+\gamma_2\right)t\right)-\frac{1}{2}\exp\left(-\gamma_1t\right) \nonumber \\
&\quad -\frac{1}{2}\exp\left(-\gamma_2t\right) + 1 \label{eq:sevth}
\end{align}
where $\chi_k=-\frac{1}{2}\gamma_kt-i\left(4g-4\omega_k\right)t$, $\varrho_k=\frac{1}{4\left(4g-i\gamma_k\right)}$, $k\in\left\{1,2\right\}$. Based on Eqs. (\ref{eq:thi}), (\ref{eq:sixth}), and (\ref{eq:sevth}), one can obtain $\mathcal{F}_g $. The numerical results, depicting the evolution of $\mathcal{F}_g $ over time $t$, are shown in Fig. \ref{Fig.1}. From Fig. \ref{Fig.1}, it can be observed that for the freely evolving system, $\mathcal{F}_g $ shows a gradual increase over time until it reaches its peak, after which it starts to decrease. This implies that the uncertainty introduced by noise affects the parameter characteristics of system (\ref{eq:sev}). Specifically, in the studied open quantum system, the coherence of system (\ref{eq:sev}) gradually becomes weak due to the influence of the environment, ultimately leading to a significant decrease in parameter estimation precision after $t=41.6$ for $g^\ast=0.1$ and $t=37.6$ for $g^\ast=0.2$. 

\subsection{\label{sec:level3.2}Feedback control}

In this subsection, we examine the impact of quantum-jump feedback control on the dynamics and analyze the control performance. Under feedback control, the system is described by Eq. (\ref{eq:twe}), and the final parameter estimation precision depends on the choice of feedback operator $U_{fb}$. Once a transition event represented by $\sigma_-^{\left(i\right)}\rho\sigma_+^{\left(i\right)}$ is detected, the operator $U_{fb}=\mathrm{exp}\left[iH_{fb}\delta_t\right]$ immediately acts on the system within an extremely short time $\delta_t$, where $H_{fb}$ is a Hermitian operator. Under the constraint of $U_{fb}U_{fb}^\dagger=I$, a feedback mechanism that can break the exchange symmetry between atoms is considered, aiming to limit destructive interference. Here, we apply feedback to only one of the qubits, i.e., local feedback. The selected feedback operator $U_{fb}$ is expressed as~\cite{29}
\begin{equation}
U_{fb}=e^{i\lambda\sigma_x}\otimes I
\label{eq:eigth}
\end{equation}
where $\lambda$ represents the feedback strength $(0\le\lambda\le\pi\ )$.

In practice, due to the non-negligible error limit of the experimental control, the efficiency $\eta$ of the detector $D$ is difficult to reach 1. The detection information directly affects the effectiveness of feedback control. Thus the analysis of detection efficiency is useful. 

\subsubsection{\textbf{Perfect detection} \texorpdfstring {$\bm{(\eta=1)}$}{}}

\begin{figure}
    \centering
    \includegraphics[width=\linewidth, height=6cm]{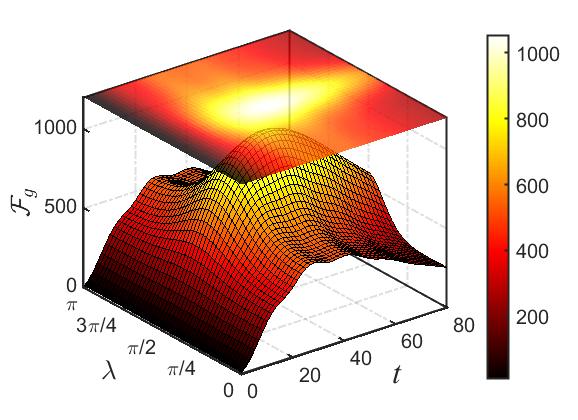}
    \caption{\label{Fig.2} The surface plot and its top view mapping of $\mathcal{F}_g$ with respect to both time $t$ and feedback strength $\lambda$ under feedback control. The true value of $g$ is set to  $g^\ast=0.1$. Other parameters are chosen as $\omega_1=\omega_2=1$, $\gamma=0.05$.}
\end{figure}

First, we analyze the case of perfect detection efficiency $\left(\eta=1\right)$, i.e., the detector successfully detects all photons and responds accordingly. In this scenario, the master equation of the system is Eq. (\ref{eq:twe}). A surface graph is plotted in Fig. \ref{Fig.2}, illustrating how $\mathcal{F}_g$ changes with varying time and feedback coefficients. For easy observation, we project this surface onto a plane. From Fig. \ref{Fig.2}, it can be observed that different feedback strengths $\lambda$ have varying effects on the function value $\mathcal{F}_g$. When $\lambda=0$, it corresponds to the scenario without feedback, and the simulation result is consistent with that of the aforementioned free evolution. As $\lambda$ changes, the surface exhibits a hump-shaped pattern, reaching  the highest point at $\lambda=\frac{\pi}{2}$. This indicates that the feedback operator $U_{fb}=e^{i\frac{\pi}{2}\sigma_x}\otimes I$ can maximize the estimation precision of the parameter, providing an optimization strategy for this scenario. Additionally, the mapped surface demonstrates that feedback has the ability to slow down the attenuation of QFI and maintain high estimation precision even over longer evolution times. 

Next, we analyze the maximum QFI that can be achieved under different control laws. Define the peak value of QFI $(\mathcal{F}_g^{max})$ and its improvement value $(\Delta\mathcal{F}_g^{max})$ as
\begin{equation}
\mathcal{F}_g^{max}=max_t\left(\mathcal{F}_g\left(t\right)\right)
\label{eq:ninth}
\end{equation}
\begin{equation}
\Delta\mathcal{F}_g^{max}=\mathcal{F}_g^{max}\left(\lambda\right)-\mathcal{F}_g^{max}\left(0\right)
\label{eq:tweth}
\end{equation}
where $\lambda=0$ corresponds to the case without control. 

\begin{figure}
    \centering
    \subfigure[]{\includegraphics[width=\linewidth, height=6cm]{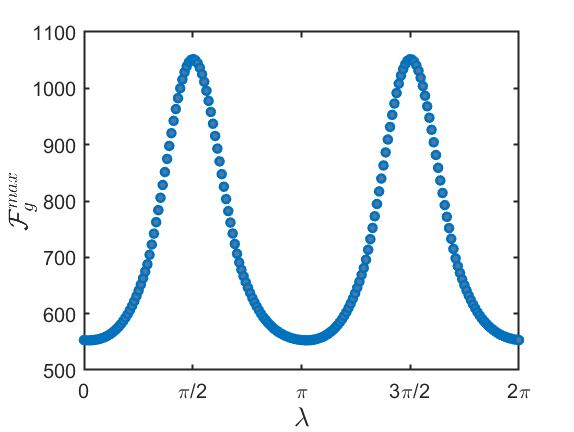}\label{Fig.3a}}
    
    \subfigure[]{\includegraphics[width=\linewidth, height=6cm]{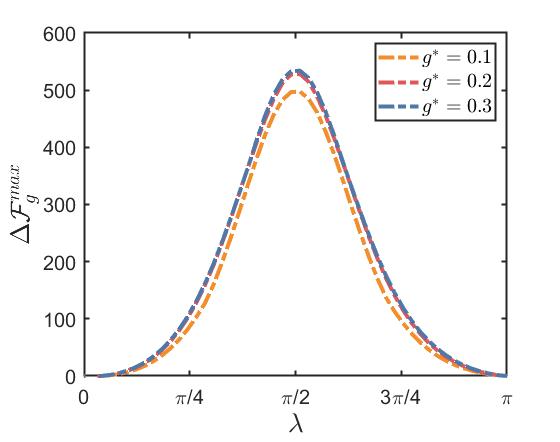}\label{Fig.3b}}
    \centering
    \caption{\label{Fig.3} (a) Under feedback control, the scatter plot of the variation of the QFI peak value $\mathcal{F}_g^{max}$ with the feedback strength $\lambda$.  The true value of $g$ is set to  $g^\ast=0.1$. (b) Under different $g^\ast$, the impact of feedback control on the QFI peak improvement degree $\Delta\mathcal{F}_g^{max}$, where the orange dash-dotted, red dash-dotted, and blue dash-dotted curves represent $g^\ast=0.1$, $g^\ast=0.2$, and $g^\ast=0.3$, respectively. Other parameters are chosen as $\omega_1=\omega_2=1$, $\gamma=0.05$.}
\end{figure}

In Fig. \ref{Fig.3}, we plot the curves of  $\mathcal{F}_g^{max}$ and $\Delta\mathcal{F}_g^{max}$ under different $\lambda$. As can be seen from Fig. \ref{Fig.3a}, $\mathcal{F}_g^{max}$ changes periodically with $\lambda$, and the period is $\pi$. Moreover, as $\lambda$ approaches $\frac{n\pi}{2}$ in the $n$th period, the degree of precision enhancement increases. At the peak value, the improvement reaches 90.18\%, which provides a valuable basis for the selection of control schemes for subsequent design of parameter recovery protocol. Considering the potential limitation of insufficient prior knowledge in practical experiments, in Fig. \ref{Fig.3b}, we illustrate the impact of feedback control on estimation precision under different $g^\ast$. Fig. \ref{Fig.3b} reveals that the three different $g^\ast$ exhibit a similar precision improvement, implying the broad applicability of this optimization strategy. In addition, we also find that as the value of $g^\ast$ increases, there is an increase in the enhancement of QFI.


\subsubsection{\textbf{Imperfect detection} \texorpdfstring {$\bm{(0<\eta<1)}$}{}}

Now consider another situation, that is, during the system transitions, some of the emitted photons are not detected by the detector $D$. 

According to the general theory of quantum measurement, the detection efficiency $\eta$ can be reflected in the measurement process, meaning that $\rho\left(t+\mathrm{d}t\right)={\eta U}_{fb}\Omega_1\left(\mathrm{d}t\right)\rho\left(t\right)\Omega_1^\dagger\left(\mathrm{d}t\right)U_{fb}^\dagger+\left(1-\eta\right)\Omega_1\left(\mathrm{d}t\right)\rho\left(t\right)\Omega_1^\dagger\left(\mathrm{d}t\right)+\Omega_0\left(\mathrm{d}t\right)\rho\left(t\right)\Omega_0^\dagger\left(\mathrm{d}t\right)(0\le\eta\le1)$, where the system performs photon detection with an efficiency of $\eta$. Correspondingly, the master equation (\ref{eq:twe}) is modified as
\begin{equation}
\begin{split} 
\dot{\rho}=&-i\left[H,\rho\right]+\eta\gamma_1\mathcal{D}\left[U_{fb}\sigma_-^{\left(1\right)}\right]\rho+ \\
&\left(1-\eta\right)\gamma_1\mathcal{D}\left[\sigma_-^{\left(1\right)}\right]\rho+\gamma_2\mathcal{D}\left[\sigma_-^{\left(2\right)}\right]\rho
\label{eq:twone}
\end{split} 
\end{equation}
with
\begin{equation}
\mathcal{D}\left[\sigma_-^{\left(k\right)}\right]\rho\!=\!\sigma_-^{\left(k\right)}\rho\sigma_+^{\left(k\right)}-\frac{1}{2}\left(\sigma_+^{\left(k\right)}\sigma_-^{\left(k\right)}\rho+\rho\sigma_+^{\left(k\right)}\sigma_-^{\left(k\right)}\right).
\label{eq:twtwo}
\end{equation}
When $\eta$ is 0, it represents a photon detection rate of 0, meaning that the feedback operator is never triggered at any time. In this case, Eq. (\ref{eq:twone}) is equivalent to Eq. (\ref{eq:sev}). When $\eta$ is 1, it corresponds to perfect detection efficiency, and Eq. (\ref{eq:twone}) reverts to Eq. (\ref{eq:twe}). When $0<\eta<1$, it corresponds to the situation that the photon detector $D$ only detects part of the photons, which is also the focus of our study in this subsection. We still use the optimal feedback operator obtained in the previous subsection, i.e., $U_{fb}=e^{i\frac{\pi}{2}\sigma_x}\otimes I$. 

\begin{figure}
    \centering
    \subfigure[]{\includegraphics[width=\linewidth, height=6cm]{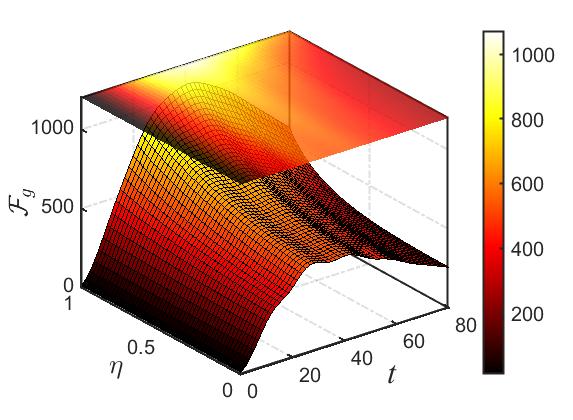}\label{Fig.4a}}
    
    \subfigure[]{\includegraphics[width=\linewidth, height=6cm]{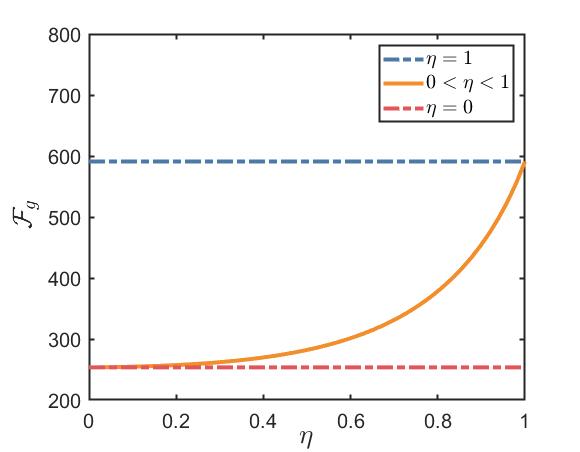}\label{Fig.4b}}
    \centering
    \caption{\label{Fig.4} (a) Under feedback control of imperfect detection, the change surface and top mapping plane of $\mathcal{F}_g$ with respect to time $t$ and detection efficiency $\eta$, where the feedback strength is set as $\lambda=\frac{\pi}{2}$. (b) When $t=80$, the change curves of $\mathcal{F}_g$ under different detection efficiencies $\eta$, where the blue dash-dotted, red dash-dotted, and orange solid curves represent $\eta=1$, $\eta=0$ and $0<\eta<1$, respectively. The true value of $g$ is set to  $g^\ast=0.1$. Other parameters are chosen as $\omega_1=\omega_2=1$, $\gamma=0.05$.}
\end{figure}

Fig. \ref{Fig.4a} shows how the QFI value $\mathcal{F}_g$ changes with varying time $t$ and detection efficiency $\eta$. Along the time axis, the objective function $\mathcal{F}_g$ generally increases with increasing detection efficiency. Even with imperfect detection efficiency, the performance under feedback control is still superior to that of free evolution, and this also can be verified more intuitively by Fig. \ref{Fig.4b}. The control effect at $t=80$ is depicted in Fig. \ref{Fig.4b}, showing that the detection efficiency curve of $\eta\in\left(0,\ 1\right)$ is between perfect detection and free evolution lines. In addition, it shows a higher detection efficiency leads to a faster rate of precision improvement. This indicates that feedback control significantly enhances the system's ability to suppress decoherence.

\subsection{\label{sec:level3.3}Hybrid control}

In the previous subsection, we have mainly analyzed the impact of quantum-jump feedback on the parameter estimation precision. The results show that this optimization strategy improves the maximum achievable QFI to some extent and slows down the rate of estimation precision decay. However, there is still room for improvement. In this subsection, Hamiltonian control is applied to further enhance the precision of extracting information of $g$. For simplicity, assuming that perfect detection efficiency can be achieved, i.e., $\eta=1$. Controls are applied in all three directions for each qubit, i.e., $H_c=\sum_{k=1,2}\sum_{j=1,2,3} u_j^{\left(k\right)}\cdot\sigma_j^{\left(k\right)}$. In this case, the total Hamiltonian can be expressed as the superposition of the free Hamiltonian and the control Hamiltonian:
\begin{equation}
H_{tot}\!=\!H_0+H_c\!=\!\omega_1\sigma_z^{\left(1\right)}+\omega_2\sigma_z^{\left(2\right)}+g\sigma_z^{\left(1\right)}\sigma_z^{\left(2\right)}+\!\!\sum_{k=1,2}\!\!{{\vec{u}}^{\left(k\right)}\!\cdot\!{\vec{\sigma}}^{\left(k\right)}}
\label{eq:twthr}
\end{equation}
where ${\vec{u}}^{\left(k\right)}$ represents the control field in all directions of the $k$th qubit. Substitute Eq. (\ref{eq:twthr}) into Eq. (\ref{eq:twe}), and the feedback operator is still chosen as $U_{fb}=e^{i\frac{\pi}{2}\sigma_x}\otimes I$. The GRAPE algorithm~\cite{14,22} is used to design the control laws. The design and analysis of Hybrid control is stated as follows.

\begin{figure}
    \centering
    \subfigure[]{\includegraphics[width=\linewidth, height=6cm]{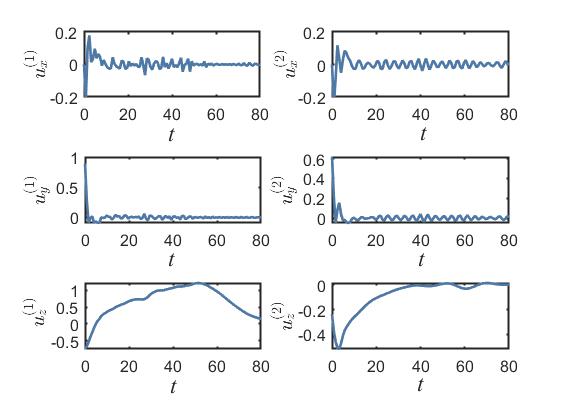}\label{Fig.5a}}
    
    \subfigure[]{\includegraphics[width=\linewidth, height=6cm]{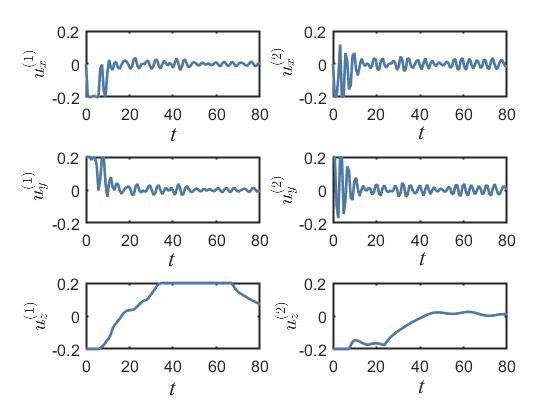}\label{Fig.5b}}
    \centering
    \caption{\label{Fig.5}(a) The optimal control law under hybrid control when there is no limit on the control amplitude. (b) Limit the control amplitude to $[-0.2, 0.2]$, the final optimal control law. In each graph, the first vertical column represents the local controls applied in three directions to the first qubit; the second vertical column represents the local controls applied in three directions to the second qubit. Other parameters are chosen as $g^\ast=0.1$, $\omega_1=\omega_2=1$, $\gamma=0.05$.}
\end{figure}

First, the target measurement time $T$ is divided into $\mathcal{M}$ segments of length $\delta t$ $\left(\delta t=\frac{T}{\mathcal{M}}\right)$. The density matrix at time $T$ can be expressed as $\rho\left(T\right)=\Pi_{n=1}^\mathcal{M}\mathrm{exp}\left(\mathcal{L}_n\delta t\right)\rho\left(0\right)$, where $\mathcal{L}_n$ represents the system evolution superoperator corresponding to the $n$th time segment, i.e., $\mathcal{L}_n\left ( \cdot \right ) =-\frac{i}{\hbar}\left [ H_0+ \sum_{k=1,2}{{\vec{u}}^{\left(k\right)}(n)\cdot{\vec{\sigma}}^{\left(k\right)}},\cdot \right ] +\mathcal{L}_\mathcal{D}\left ( \cdot \right ) $. According to Eq. (\ref{eq:thi}), we calculate the gradient of the objective function at time $T$:
\begin{equation}
\frac{\partial\mathcal{F}_g\left(T\right)}{\partial u_j^{\left(k\right)}\left(m\right)}\!=\!\mathrm{Tr}\!\left(\!\frac{\partial\mathcal{N}_m}{\partial u_j^{\left(k\right)}\left(m\right)}\rho\left(m\delta t\right)\!\right)\!+\!\mathrm{Tr}\!\left(\!\mathcal{N}_m\frac{\partial\rho\left(m\delta t\right)}{\partial u_j^{\left(k\right)}\left(m\right)}\!\right)
\label{eq:twfor}
\end{equation}
where $u_j^{\left(k\right)}\left(m\right)$ represents the control law for the $m$th time segment, and $\mathcal{N}_m=\Pi_{n=m+1}^\mathcal{M}\mathrm{exp}\left(\mathcal{L}_n\delta t\right)\left[L_s^\theta\left(T\right)\right]^2$. Then, the control variables are updated according to the gradient information:
\begin{equation}
u_j^{\left(k\right)}\rightarrow u_j^{\left(k\right)}+\epsilon\frac{\partial\mathcal{F}_g\left(T\right)}{\partial u_j^{\left(k\right)}}
\label{eq:twfif}
\end{equation}
where $\epsilon$ is the learning rate. We set it as $\epsilon=0.01$, the measurement time as $T=80$, the number of time segments as $\mathcal{M}=100$, and the number of iterations as 500 times. Fig. \ref{Fig.5a} shows how $u_j^{\left(k\right)}$ changes with the time $t$ in six directions when the control amplitude is unconstrained after iterations. Considering some physical and technical constraints, we may need to limit the control amplitude within a certain range. From Fig. \ref{Fig.5a}, we can see that the optimal control amplitudes in different directions differ by more than 5 times. In addition, during most of the time, the control values are within the range of $\pm0.2$. This indicates that the control amplitude of the hybrid control scheme can be limited to $[-0.2, 0.2]$. The result after imposing restrictions on the control amplitude is shown in Fig. \ref{Fig.5b}.

\begin{figure}
    \centering
    \subfigure[]{\includegraphics[width=\linewidth, height=6cm]{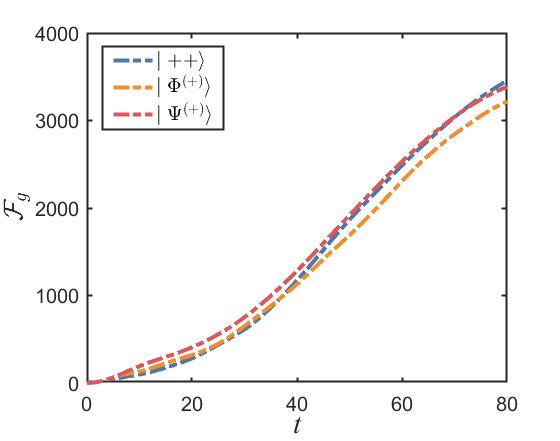}\label{Fig.6a}}
    
    \subfigure[]{\includegraphics[width=\linewidth, height=6cm]{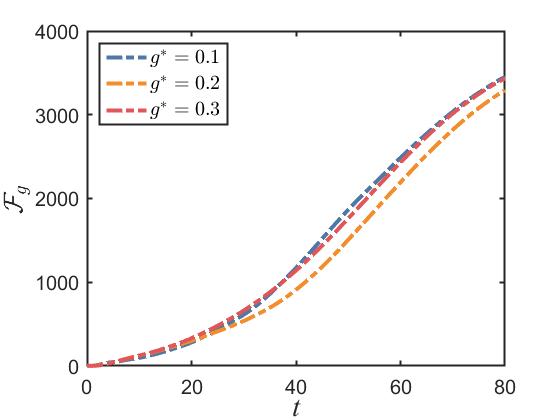}\label{Fig.6b}}
    \centering
    \caption{\label{Fig.6} (a) Under the hybrid control scheme, the influence of different initial probe states on the precision of unknown parameter estimation, where the blue dash-dotted, orange dash-dotted, and red dash-dotted curves represent $\ket{++}$, $\ket{\Phi^{(+)}}$ and $\ket{\Psi^{(+)}}$, respectively. (b) When the parameter to be estimated takes different true values, the curves of $\mathcal{F}_g$ with respect to time $t$, where the blue dash-dotted, orange dash-dotted, and red dash-dotted curves represent $g^\ast=0.1$, $g^\ast=0.2$ and $g^\ast=0.3$, respectively. Other parameters are chosen as $\omega_1=\omega_2=1$, $\gamma=0.05$.}
\end{figure}

Next we discuss the dependence of the hybrid control scheme on the initial probe state. We choose three typical quantum states for comparison: $\ket{++}$, $\ket{\Phi^{(+)}}=\left(\ket{00}+\ket{11}\right)/\sqrt2$, and $\ket{\Psi^{(+)}}=\left(\ket{01}+\ket{10}\right)/\sqrt2$. The QFI values $\mathcal{F}_g$ in these three states are plotted in Fig. \ref{Fig.6a}, revealing the similar upward trend. It indicates that the hybrid control scheme can quickly adjust the initial state to the state which is sensitive to the unknown parameter. Further, it implies that the system exhibits a high degree of freedom in the selection of the initial probe state, not limited to specific quantum states. In addition, the change curves of $\mathcal{F}_g$ under different $g^\ast$ are plotted in Fig. \ref{Fig.6b}. It can be seen that with the increase of evolution time, the QFI of the three scenarios has been greatly improved, which reflects good robustness of the algorithm. At the same time, it indicates that the hybrid control scheme has a low dependency on prior knowledge and is more suitable for practical scenarios.

\begin{figure}
    \centering
    \subfigure[]{\includegraphics[width=\linewidth, height=6cm]{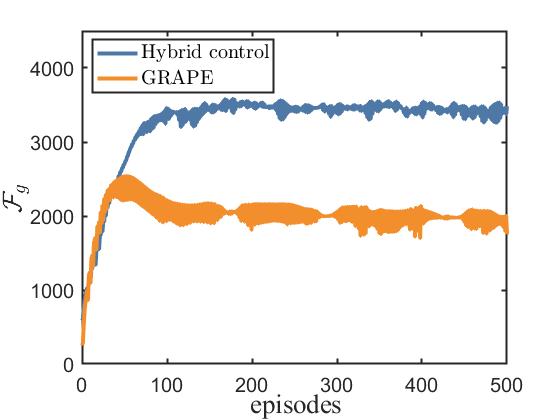}\label{Fig.7a}}
    
    \subfigure[]{\includegraphics[width=\linewidth, height=6cm]{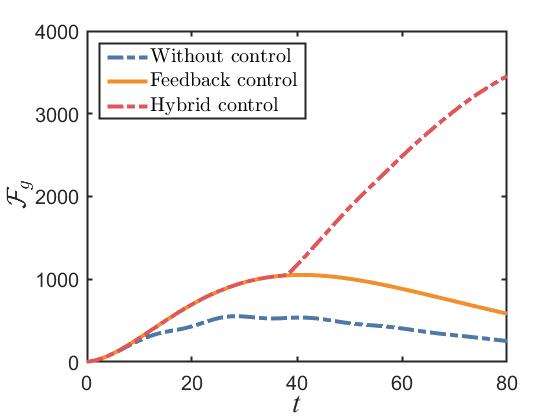}\label{Fig.7b}}
    \centering
    \caption{\label{Fig.7} (a) After 500 iterations, a comparison of the improvement in estimation precision between the hybrid control scheme and the GRAPE algorithm. The blue solid curve represents the case where feedback control and additional Hamiltonian control are simultaneously applied to the system, and the orange solid curve represents the case where only the GRAPE algorithm is used to apply Hamiltonian control. (b) The variation curves of $\mathcal{F}_g$  with respect to time $t$ for three scenarios, where the blue dash-dotted, orange solid, red dash-dotted curves represent no control applied, only feedback control applied, and the final hybrid control applied, respectively. Other parameters are chosen as $g^\ast=0.1$, $\omega_1=\omega_2=1$, $\gamma=0.05$.}
\end{figure}

Finally, for making a comparison with the traditional GRAPE algorithm,  in Fig. \ref{Fig.7a}, we plot the curves of $\mathcal{F}_g$ under hybrid control scheme and GRAPE algorithm during the iteration process. Due to the correction effect of feedback control in the early stage, the hybrid scheme shows superior performance. After the 67th iteration, the performance is much higher than the traditional GRAPE algorithm, which once again verifies the necessity of feedback control. Subsequently, both curves fluctuate around their respective stable values. Note that during the iteration process, the two curves do not always exhibit the upward trend. This is due to the non-concave nature of the objective function, which can lead the GRAPE algorithm to get stuck in local optimum while searching for control pulses. Fig. \ref{Fig.7b} intuitively shows the curves of $\mathcal{F}_g$ under three different cases: no control, feedback control, and hybrid control. Before $T=40$, additional Hamiltonian control has little impact on precision improvement, and feedback control plays a key role. For simplicity, we choose the feedback control curve instead. As the decoherence effect gradually becomes apparent, the overall estimation precision without control shows the decreasing trend. However, appropriate feedback control can suppress this decreasing trend. After that, the improvement effect of hybrid control on the QFI value gradually becomes evident, and eventually reaching 6.24 times of the maximum precision value of free evolution. The above simulation results show that even after the long evolution, hybrid control can still maintain distinguishability between quantum trajectories corresponding to different parameter true values, and can extend the coherence time of the system. 

\section{\label{sec:level4}Bayesian Batch Single-Parameter Quantum Recovery Protocol}

In this section, we study the other two steps of parameter estimation, namely measurement selection and parameter recovery. Further combining these steps with the hybrid control scheme discussed in Section \ref{sec:level3}, an adaptive Bayesian batch single-parameter quantum recovery protocol is proposed. 

Before proceeding with protocol design, we first introduce the Bayesian estimation theory needed for parameter recovery. For the majority of problems that use prior knowledge to deal with uncertainty, Bayesian estimation demonstrates good performance. It can provide a complete probability distribution and a unique state estimate, which has wide applications in parameter estimation, decision analysis, and other fields~\cite{32}. In quantum parameter estimation task, for the parameter $\theta$ to be estimated, it can use a set of specific POVM $\left\{\mathrm{M}_y\right\}$ to perform the same measurement operation on $N$ copies of $\rho(\theta)$. This can transform the quantum estimation problem into a classical statistical probability problem. Based on the observed outcome $y$, the prior probability is defined as $P\left(y\middle|\theta\right)$, and the posterior probability $P\left(\theta\middle| y\right)$ is updated after each measurement. The Bayesian rule can be expressed as
 \begin{equation}
P\left(\theta\middle| y\right)=\frac{P\left(y\middle|\theta\right)P\left(\theta\right)}{\int{P\left(y\middle|\theta\right)P\left(\theta\right)\mathrm{d} }\theta}.
\label{eq:twsix}
\end{equation}

Correspondingly, the Bayesian estimator can be expressed as
 \begin{equation}
\hat{\theta}=\int{\theta P\left(\theta\middle| y\right)\mathrm{d} }\theta.
\label{eq:twsev}
\end{equation}

In addition to the classical probability and statistical estimation rules, a quantum parameter recovery protocol also encompasses measurement rules and simulation design~\cite{32}, which will be discussed in the following subsections.

\subsection{\label{sec:level4.1}Adaptive measurement rules}

In Section \ref{sec:level3}, the selection of the optimal measurement was not considered in the control scheme. The reason is that the lower bound of the variance achievable by all physically allowed measurement operators is equivalent to the reciprocal of QFI. However, when it comes to a practical recovery protocol, we must consider an actual measurement strategy, which is directly related to the extraction of quantum parameter information and significantly affects the final estimation precision. In cases where the prior knowledge is abundant, it is possible to directly obtain better measurements that make CFI close to QFI. Here, our main focus lies in the situation of limited prior knowledge, which is more frequently encountered in practical applications. To address this, we propose an adaptive scheme that combines Bayesian estimation. This scheme continuously acquires new estimated value based on the observed data and then use the new estimated value to update the measurement operators. The specific measurement rules are stated as follows.

The first step is the selection of an initial set of measurements with good robustness, which has high sensitivity to a large range of parameters to be estimated. Consider the estimation problem of the coupling coefficient $g$ in the ZZ-coupled system. By calculating the optimal measurements for $g^\ast=0.1$, and approximating it, the POVM set is obtained as follows:
\begin{equation}
\begin{aligned}
\mathrm{M}_0&=\frac{1}{4}\begin{bmatrix}
  1&  i&  i& 1\\
  -i&  1&  1& -i\\
  -i&  1&  1& -i\\
  1&  i&  i& 1
\end{bmatrix} \\
\mathrm{M}_1&=\frac{1}{4}\begin{bmatrix}
  1&  i&  -i& -1\\
  -i&  1&  -1& i\\
  i&  -1&  1& -i\\
  -1&  -i&  i& 1
\end{bmatrix} \\
\mathrm{M}_2&=\frac{1}{4}\begin{bmatrix}
  1&  -i&  i& -1\\
  i&  1&  -1& -i\\
  -i&  -1&  1& i\\
  -1&  i&  -i& 1
\end{bmatrix} \\
\mathrm{M}_3&=\frac{1}{4}\begin{bmatrix}
  1&  -i&  -i& 1\\
  i&  1&  1& i\\
  i&  1&  1& i\\
  1&  -i&  -i& 1
\end{bmatrix}.
\end{aligned}
\label{eq:tweig}
\end{equation}

Multiple simulations have shown that this set of POVM exhibits a high distinguishability for different values of $g$ within a large range, that is, the measurement results vary significantly. Therefore, Eq.(\ref{eq:tweig}) is selected as the set of initial measurements for this scenario. 

After obtaining the initial measurements, the practical measurement overhead problem needs to be considered from the perspective of adaptive correction measurement. Based on the parameter information obtained from the early measured simulation data, new estimated value can be obtained. By solving Eq. (\ref{eq:for}), we obtain the SLD operator $L_s^\theta$ corresponding to the estimated value. Then a new set of projective measurement bases $E_i$ is obtained. Note that updating the POVM after each measurement result can achieve high precision quickly, but it significantly increase the computational cost and time overhead. To solve this problem, we batch the measurement results to approximate the single update, i.e., applying the same set of measurements to a batch of $N$ replicas, and then performing a round of updates based on the full probability distribution obtained by Bayesian estimation.

\subsection{\label{sec:level4.2}Simulation design}

Combining with the aforementioned adaptive measurement rules, the numerical simulation is conducted to  illustrate the parameter recovery protocol. We still consider the parameter estimation problem of the coupling coefficient $g$ in a ZZ-coupled system. In the simulation, the parameters are set as follows. Take 100 values of the coupling coefficient $g$ at equal intervals within the range of $\left[0,\ 0.2\right]$, and the initial probability distribution of $g$ is uniform, i.e., $P_0\left(g\right)=\frac{1}{100}$, where $g\in\left[0,\ 0.2\right]$. The initial probe state is chosen as $\ket{++}$, and the set of initial measurements follows Eq. (\ref{eq:tweig}). The true value of $g$ is set as $g^\ast=0.1$. The target measurement time is $T=80$, and the total evolution time is divided into $\mathcal{M}=100$ equal parts. For simplicity, we assume the photodetector efficiency $\eta$ in feedback control to be $1$. The simulation design includes the following steps.

First, we simulate the real evolution process of a ZZ-coupled system. The density matrix $\rho_g\left(T\right)$ corresponding to each $g$ at the target time $T$ is obtained by placing the initial probe state in the parameter channel and allowing it to evolve. When $g=0.1$, we mark its corresponding density matrix as the real matrix $\rho_{g^*}\left(T\right)$. In this case, we simulate the actual evolution of the quantum state. Then, calculating the probabilities of four measurement outcomes $\left\{m_0,m_1,m_2,m_3\right\}$ based on $P^\ast\left(m_l\right)=\mathrm{Tr}\left[\rho_{g^*} \left(T\right)\mathrm{M}_l\right]$, in order to generate a sample sequence. 

Next, we need to obtain the sequences of measurement results by sampling. Here, the entire sampling process is divided into $N=20$ batches, with each batch consisting of repeated simulations with $R=100$ copies. Within each batch, the same set of measurements is used, resulting in a sequence $\vec{\mathcal{S}}$ of length 100 in each batch, where each element of the sequence corresponds to one of the four measurement outcomes. Considering the possible projection measurement errors in practice and the limited number of simulations, we discuss two scenarios: perfect sample ${\vec{\mathcal{S}}}_p$ and imperfect sample ${\vec{\mathcal{S}}}_{np}$. Perfect sample ${\vec{\mathcal{S}}}_p$ implies that measurement results are generated strictly according to the theoretical probabilities with equal proportions, and the required sample sequence is generated by randomly shuffling the order. Imperfect sample ${\vec{\mathcal{S}}}_{np}$ is generated using the roulette wheel method, where the cumulative probability of each measurement result is calculated as
 \begin{equation}
d_l=\sum_{n=0}^{l}{P^\ast\left(m_n\right)},l\in\left\{0,1,2,3\right\}.
\label{eq:twnin}
\end{equation}

\begin{figure*}
    \centering
    \includegraphics[width=\linewidth, height=9cm]{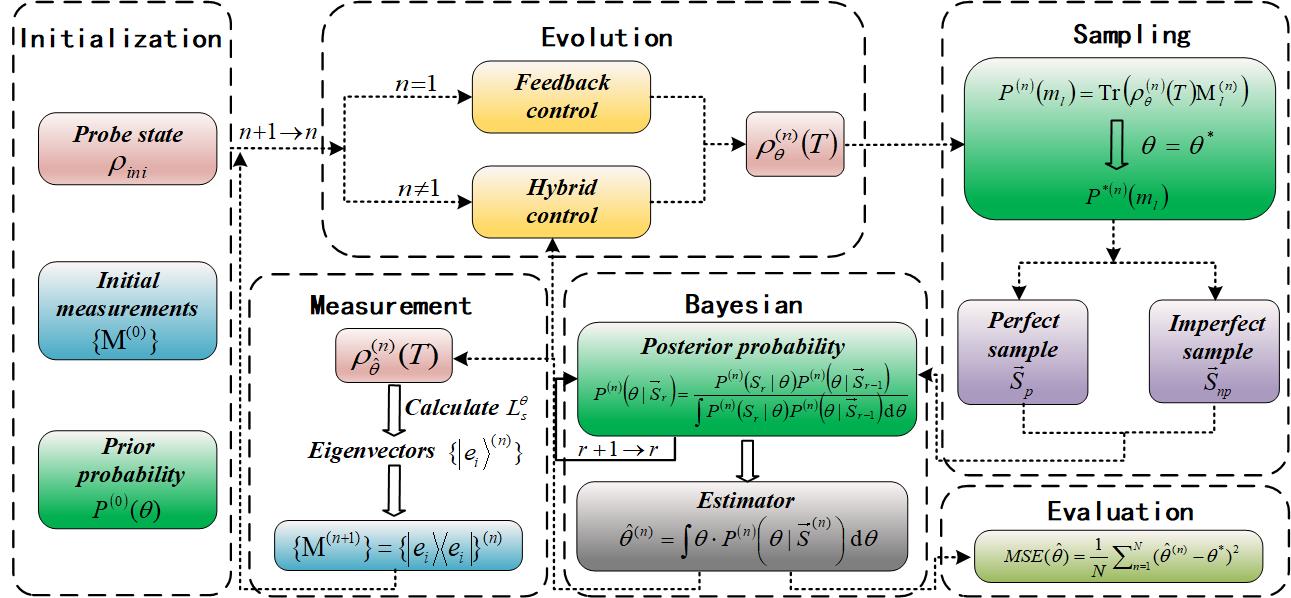}
    \caption{\label{Fig.8} The framework of the single-parameter quantum recovery protocol proposed in this paper. The protocol includes initialization, evolution, sampling, Bayesian estimation, adaptive measurement, and evaluation steps.}
\end{figure*}

A computer generates a random number in $\left(0,\ 1\right)$. The corresponding measurement result depends on the interval it falls into. For example, if the random number is in $(0,\ d_0]$, the measurement result is $m_0$; if the random number is in $(d_{l-1},\ d_l]$, the measurement result is $m_l$. Repeat the above process 100 times to obtain the sample sequence. 

Then, based on the sample sequence, Bayesian estimation theory is employed to calculate the result probability for  $g$ with different values, denoted as $P(\mathcal{S}_r|g)$. Then, the probability distribution of $g$ is updated sequentially according to the probability update formula:
 \begin{equation}
P\left(g\middle|{\vec{\mathcal{S}}}_r\right)=\frac{P\left(\mathcal{S}_r\middle| g\right)P\left(g|{\vec{\mathcal{S}}}_{r-1}\right)}{\int{P\left(\mathcal{S}_r\middle| g\right)P\left(g|{\vec{\mathcal{S}}}_{r-1}\right)\mathrm{d} }g}
\label{eq:thity}
\end{equation}
where $\mathcal{S}_r$ represents the $r$th result of the measurement result sequence $\vec{\mathcal{S}}$.

According to the total probability formula, after completing the iterations for all samples in the current batch, the Bayesian estimator can be used to obtain the new estimated value $\hat{g}=\int{gP\left(g\middle|{\vec{\mathcal{S}}}\right)\mathrm{d} }g$. Meanwhile, the POVM for the next batch can be determined based on $\hat{g}$ and the eigenvectors of the SLD as described in Eq. (\ref{eq:for}). Proceeding with iterative batches in this manner until the measurements of all batches are completed. Throughout this process, the estimated value of each batch is recorded to observe the estimation effect. 

Finally, we determine an evaluation indicator for the parameter recovery results. Here, the precision of the estimated value is evaluated using the mean square error (MSE) of the estimated value:
 \begin{equation}
MSE\left(\hat{g}\right)=\frac{1}{N}\sum_{n=1}^{N}\left({\hat{g}}^{(n)}-g^\ast\right)^2
\label{eq:thione}
\end{equation}
where ${\hat{g}}^{(n)}$ represents the $n$th batch of estimated values. Note that due to factors such as experimental overhead, time cost, and limited prior knowledge, the final estimation precision is difficult to reach the quantum limit. But it has the advantages of simplicity, efficiency, and practicality in implementation. Since this protocol strikes a balance between precision requirements and practical resources, it is considered one of the best candidate strategies.

In the next subsection, we integrate the two control schemes proposed in Section \ref{sec:level3} into the parameter recovery protocol and analyze the simulation performance. The entire parameter recovery protocol flowchart is shown in Fig. \ref{Fig.8}.

\subsection{\label{sec:level4.3}Performance analysis}

\begin{figure}
    \centering
    \includegraphics[width=\linewidth, height=6cm]{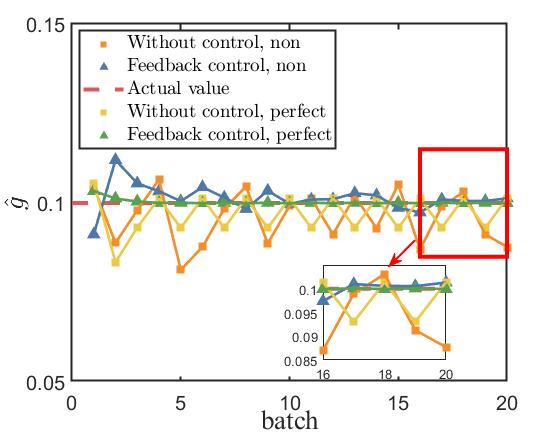}
    \caption{\label{Fig.9} The estimated value $ \hat{g}$ for the ZZ-coupled system with different sample levels over iterations, with and without feedback control. Additionally, an enlarged view is provided for the results of batches 16 to 20. The red dotted curve represents the true value of $g$, i.e., $g^\ast=0.1$; the orange and yellow squares represent the estimation results of imperfect sample and perfect sample without control, respectively. The blue and green triangles represent the estimation results of imperfect sample and perfect sample under feedback control, respectively. Other parameters are chosen as $\omega_1=\omega_2=1$, $\gamma=0.05$.}
\end{figure}

First, we discuss the effect of feedback control in parameter recovery protocol. The feedback operator is still chosen as $U_{fb}=e^{i\frac{\pi}{2}\sigma_x}\otimes I$. Using two different levels of samples, the trend of estimation value is evaluated before and after applying feedback control, as shown in Fig. \ref{Fig.9}. It can be seen that under free evolution, the estimated values exhibit significant fluctuations around the true value. The enlarged graph more intuitively shows that there is no decreasing trend in fluctuations without control even at the end of the iterations. In contrast, the estimation values under feedback control, whether with perfect or imperfect samples, show a better convergence trend. In the case of perfect samples, the estimated values closely match the true value in early iterations. This improvement trend in estimation precision is consistent with that under the previous schemes in Section \ref{sec:level3}. 

Next, we discuss the effect of hybrid control in parameter recovery protocol. Additional Hamiltonian control is an optimization control and can easily lead to local optimum or failure to converge in the absence of prior knowledge. Thus we choose to apply only feedback control in the simulations of batch $n=1$ to measure some of the parameter information. Upon entering batch $n=2$, the hybrid scheme is initiated. The GRAPE algorithm is utilized to obtain a set of control laws $\sum_{k=1,2}{\vec{u}}^{\left(k\right)}$, which is used as the Hamiltonian control in the next batch and also as the initial control of the GRAPE algorithm. This process iterates sequentially, and the update rules for measurements remain consistent with that of feedback control. Finally, the MSE for different scenarios are presented in Table \ref{tab:t1}. It can be observed that under the same control scheme, perfect samples exhibit lower error values compared to imperfect samples. Regardless of the control scheme, the estimation variance is significantly lower than that under free evolution. This validates the effectiveness of combining adaptive measurement updates with control schemes, making the entire quantum parameter recovery protocol easier to implement and more efficient.


\begin{table}
\caption{\label{tab:t1}Estimated mean square error of parameter to be estimated under different scenarios}
\renewcommand\arraystretch{1.2}
\begin{ruledtabular}
\begin{tabular}{ccc}
Control scheme&Imperfect sample&Perfect sample\\
\hline
Without control & $7.04 \times 10^{-5}$ & $3.97 \times 10^{-5}$ \\
Feedback control & $3.67 \times 10^{-5}$ & $1.75 \times 10^{-5}$ \\
Hybrid control & $1.79 \times 10^{-5}$ & $5.47 \times 10^{-6}$ \\
\end{tabular}
\end{ruledtabular}
\end{table}

\section{\label{sec:level5}Conclusion}

In this paper, we have considered the overall process of quantum parameter estimation problem for a ZZ-coupled system. On the one hand, to improve the estimation precision of the coupling coefficient $g$, two control schemes, i.e., feedback control and hybrid control, are proposed. In feedback control scheme, we have found a stable feedback operator that can significantly improve estimation performance by acting on only one qubit. In hybrid control scheme, combining feedback control with additional Hamiltonian control, the estimation precision is further improved. Compared with free evolution, both control schemes exhibit superior estimation performance and significantly slow down the rate of system decoherence. On the other hand, to enhance the connection between the theoretical scheme and the practical protocol, a practical quantum parameter recovery protocol based on the Bayesian estimation method has been proposed. This protocol combines batch adaptive update measurement with two aforementioned control schemes. The parameter recovery results verify the effectiveness of schemes in improving precision. The protocol proposed in this paper is also applicable to improving other quantum performances, such as stable entropy squeezing in atomic systems. Future research includes exploring the problem of multi-parameter estimation of different quantum systems using hybrid control scheme. Under different actual precision requirements, the improvement of practical single-parameter or multi-parameter recovery protocols will also be a direction worth studying.

\begin{acknowledgments}
The authors acknowledge the support from the National Natural Science Foundation of China under Grant No. 62373342.
\end{acknowledgments}

\nocite{*}


\end{document}